\documentclass[11pt]{article}
\usepackage{mathptmx}
\usepackage{sprinconf}
\usepackage{graphicx}
\usepackage{bm}
\usepackage{longtable}

\begin{document}

\author{Elena Santopinto  and Giuseppe Galat\`a}

\affil{ INFN and Universit\`a di Genova,
via Dodecaneso 33, 16146 Genova, Italy}
\title{Spectroscopy of tetraquark mesons.}

\beginabstract

A classification of tetraquark states in terms of the spin-flavour, colour and spatial 
degrees of freedom has been constructed.
The permutational symmetry properties of both the spin-flavour and orbital parts of the
quark-quark and antiquark-antiquark subsystems are briefly discussed.
This classification is useful both for model-builders and experimentalists.
An evaluation of the tetraquark spectrum is obtained from a Iachello mass formula
for tetraquark systems. The ground state tetraquark nonet is identified with
$f_{0}(600)$, $\kappa(800)$, $f_{0}(980)$, $a_{0}(980)$.  The mass splittings predicted
 by this mass formula are in agreement with the KLOE, Fermilab E791 and BES experimental data.
\endabstract

\section{\label{sec:introduzione}Introduction}

The KLOE, E791 and BES collaborations have recently provided evidence of the low mass resonances $f_{0}(600)$ \cite{Aloisio:2002bt} \cite{Aitala:2000xu} \cite{Ablikim:2005ni} and $\kappa(800)$ \cite{Aitala:2000xu} \cite{Ablikim:2005ni} and this has triggered new interest in meson spectroscopy. Maiani et al. \cite{Maiani:2004uc} have suggested that the lowest lying scalar mesons, $f_{0}(980)$, $a_{0}(980)$, $\kappa(800)$ and 
$f_0(600)$ could be described as complex tetraquark states, in particular as a diquark and antidiquark system, since the quark-antiquark assignment to P-waves \cite{Tornqvist:1995kr} has never
 worked in the scalar case \cite{Jaffe:1976ig}. 
The $f_{0}(980)$, in a simple quark-antiquark scheme, is associated with non-strange quarks \cite{Tornqvist:1995kr} and so it is difficult to explain both its higher mass and its decay properties \cite{Jaffe:1976ig} \cite{Maiani:2004uc}.  Already in the seventies Jaffe \cite{Jaffe:1976ig} studied a four quark bag model and suggested the tetraquark structure of the scalar nonet. Other identifications have been proposed  \cite{Close:2002zu}, in particular 
as quasimolecular-states \cite{Weinstein:1982gc} and as dynamically generated resonances \cite{Oset}.
For review articles both on the experiments and on the theoretical models we refer the reader to \cite{Amsler:2004ps,Close:2002zu,PDG} and references therein. 

We address the problem of constructing a complete classification scheme of the two quark-two antiquark states in terms of SU(6)$_{~sf}$ \cite{Santopinto:2006my}.
An evaluation of the tetraquark spectrum for the lowest scalar mesons is obtained \cite{Santopinto:2006my} from a generalization of the Iachello mass formula
for normal mesons \cite{Iachello:1991fj}.

\section{\label{sec:classificazionestati}The classification of tetraquark states}

The tetraquark wave function contains contributions connected to the spatial degrees of freedom and the internal degrees of freedom of colour, flavour and spin.
We shall make use as much as possible of symmetry principles without, for the moment, introducing any explicit dynamical model.
In the construction of the classification scheme we are guided by two conditions: the tetraquark wave functions should be a colour singlet, as all physical states, and since tetraquarks are composed of two couples of identical fermions, their states must be antisymmetric for the exchange of the two quarks and the two antiquarks.

The allowed SU(3)$_{f}$ representations for the $qq\bar q\bar q$ system are obtained by means of the product $[3]\otimes [ 3]\otimes [\bar 3]\otimes [\bar 3] =[1]\oplus [8]\oplus [1]\oplus [8]\oplus [27]\oplus [8]\oplus [8]\oplus [10]\oplus [\overline{10}]$. The allowed isospin values are $I=0,\frac{1}{2},1,\frac{3}{2},2$ , while the hypercharge values are $Y=0,\pm 1,\pm 2$. The values $I=\frac{3}{2},2$ and $Y=\pm 2$ are exotic, which means that they are forbidden for the $q\bar q$ mesons.
The allowed SU(2)$_{s}$ representations are obtained by means of the product $ [2]\otimes [2]\otimes [2]\otimes [2] =[1]\oplus [3]\oplus [1]\oplus [3]\oplus [3]\oplus [5]$. 
The tetraquarks can have an exotic spin $S=2$,value forbidden for $q\bar q$ mesons.
The SU(6)$_{sf}$-spin-flavour classification is obtained
 by $[6]\otimes [ 6]\otimes [\bar 6]\otimes [\bar 6]=[1]\oplus [35]\oplus [405]\oplus [1]\oplus [35]\oplus [189]\oplus [35]\oplus [280]\oplus [35]\oplus [\overline{280}]$. In Appendix A of Ref.\cite{Santopinto:2006my}all the flavour states in the $qq\bar q \bar q$ configuration are explicitly written. Moreover in the ideal mixing hypothesys, the flavour states of the tetraquarks are a superposition of the previous states in such a way to have defined strange quark and antiquark numbers. Clearly the only states that can be mixed are those with the same quantum numbers, i.e. same isospin and hypercharge, and they are explicitly constructed in Ref.\cite{Santopinto:2006my}. This explicit construction can be important for other applications. The ideal mixing is essentially a consequence of the OZI rule. This hypothesys remains to be proved, but it is used by all the authors  working on $q\bar q$ and tetraquarks.
Tetraquarks are made up of four objects, so we have to define three relative coordinates that we choose as \cite{Estrada}, that is the two relative coordinates between two quarks and two antiquarks and the relative coordinate between their centers of mass to which are associated three orbital angular momenta, $L_{13}$, $L_{24}$ and $L_{12-34}$ respectively.  
We have four different spins and three orbital angular momenta and the total angular momentum $J$ is obtained by combining spins and orbital momenta.
The parity for a tetraquark system is the product of the intrinsic parities of the quarks and the antiquarks times the factors coming from the spherical harmonics  \cite{Estrada} as
$P=P_{q}P_{q}P_{\bar q}P_{\bar q}(-1)^{L_{13}}(-1)^{L_{24}}(-1)^{L_{12-34}}=(-1)^{L_{13}+L_{24}+L_{12-34}}$.
Using these coordinates, the charge conjugation eigenvalues can be calculated by following the same steps as in the $q\bar q$ case, considering a tetraquark as a $Q\bar Q$ meson, where $Q$ represents the couple of quarks and $\bar Q$ the couple of antiquarks, with total \textquotedblleft spin\textquotedblright $S$ and relative angular momentum $L_{12-34}$. The $C$ eigenvectors are those states for which $Q$ and $\bar Q$ have opposite charges and the $C$ operator eigenvalues are \cite{Estrada} $C=(-1)^{L_{12-34}+S}$. The eigenvalues of the G parity are in Ref.\cite{Santopinto:2006my}.
Tetraquark mesons do not have forbidden $J^{~PC}$ combinations. 
Since tetraquarks are composed of two couples of identical fermions, their states must be antisymmetric for the exchange of the two quarks and the two antiquarks and so it is necessary to study the permutational symmetry (i.e. the irreducible representations of $S_{~2}$) of the colour, flavour, spin and spatial parts of the wave functions of each subsystem, two quarks and two antiquarks. Moreover we have another constrainth due to the fact that  
only the singlet colour states are physical states. Regarding the two colour singlets it is better to write them by underlining their permutational $S_{2}$ symmetry, antisymmetric (A) or symmetric (S): $(qq)$ in $\;[\bar 3]_{C}\;(A)\;$ and $\;(\bar q\bar q)\;$ in $ [3]_{C}\; (A)$, or  $(qq)\;$ in $ [6]_{C} (S)$ and $\; (\bar q\bar q)\;$ in $\; [\bar 6]_{C}\; (S)$. 
Then one should study the permutational symmetry of the spatial part of the two quarks (two antiquarks) states and the permutational symmetry of the SU(6)$_{~sf}$ representations for a couple of quarks. The spatial, flavour, colour and spin parts with given permutational symmetry ($S_{2}$) must then be arranged together to obtain completely antisymmetric states under the exchange of the two quarks and the two antiquarks. The resulting states are listed in Table 
III of  Ref. \cite{Santopinto:2006my}.
In Table V, VI, VII and VIII of Ref. \cite{Santopinto:2006my} the possible flavour, spin and $J^{~PC}$ for different orbital angular momenta are studied. 

In 1991 Iachello, Mukhopadhyay and Zhang developed a mass formula \cite{Iachello:1991fj} for $q\bar q$ mesons, which is a generalization of the Gell-Mann Okubo mass formula 
\begin{equation}
M^{2}=(N_{n}M_{n}+N_{s}M_{s})^{2}+a~\nu +b~L+c~ S+d~J+e~ M'^{2}_{iji'j'}+f~ M''^{2}_{iji'j'},
\label{eq:formulamassa}
\end{equation}
where $N_{n}$ is the non-strange quark and antiquark number, $M_{n}=M_{u}=M_{d}$ is the non-strange constituent quark mass, $N_{s}$ is the strange quark and antiquark number, $M_{s}$ is the strange constituent quark mass, $\nu $ is the vibrational quantum number, $L$, $S$ and $J$ are the total orbital angular momentum, the total spin and the total angular momentum respectively, $M~'^{2}_{iji'j'}$ and $M''^{2}_{iji'j'}$ are two phenomenological terms which act only on the lowest pseudoscalar mesons, the first acts on the octect and  encodes the unusually low masses of the eight Goldstone bosons, while the second acts on the $\eta $ and $\eta '$ and encodes the non-negligible $q\bar q$ annihilation effecs that arise when the lowest mesons are flavour diagonal.
Flavour states in the ideal mixing hypothesis are considered, except for the lowest pseudoscalar nonet.
During these 15 years the values reported by the PDG regarding the mesons are changed in a considerable way, so we have decided to updated the fit of the Iachello model using the values for the light $q \bar q$ mesons reported by the last PDG \cite{PDG}. The resulting values for the parameters are reported in Ref. \cite{Santopinto:2006my}. 
\begin{table}[ht]
\begin{center}
\caption{The candidate tetraquark nonet. Experimental data and quantum numbers}
\begin{tabular}{lllll}
\topline
Meson & $I^{G}(J^{PC})$ & $ N_s$ & $Mass ~(GeV)$ & Source \\
\midline
$a_{0}(980)$ & $1^{-}(0^{++})$ & 2 & $0.9847\pm 0.0012$  & PDG \cite{PDG} \\
$f_{0}(980)$ & $0^{+}(0^{++})$ & 2 & $0.980\pm 0.010$ &  PDG \cite{PDG}  \\
$f_{0}(600)$ & $0^{+}(0^{++})$ & 0 & $0.478\pm 0.024$ &  KLOE \cite{Aloisio:2002bt}  \\
$k(800)$ & $\frac{1}{2}(0^{+})$ & 1 & $0.797\pm 0.019$ &  E791 \cite{Aitala:2002kr}  \\
\bottomline
\end{tabular}
\end{center}
\end{table}
The Iachello mass formula was originally developed for $q\bar q$ mesons. In order to describe uncorrelated tetraquark systems by means of an algebraic model one should introduce a new spectrum generating algebra for the spatial part, i.e. U(10) since we have nine spatial degrees of freedom. We will not address this difficult problem, but we choose to write the part of the mass formula regarding the internal degrees of freedom in the same way. The splitting inside a given flavour multiplet, to which is also associated a given spin, can be well described by means of the part of the Iachello mass formula that depends only on the numbers of strange and non-strange quarks and antiquarks. 
In order to determine the mass splitting of the candidate tetraquark nonet, we can simply use, see Ref.\cite{Santopinto:2006my},
\begin{equation}
\label{eq:tetrascorrelati}
M^{2}=\alpha +(N_{n}M_{n}+N_{s}M_{s})^{2},
\end{equation}
where $\alpha $ is a constant that encodes all the spatial and spin dependence of the mass formula, and $M_{n}$ and $M_{s}$ are the masses of the constituent quarks as obtained from the  previously discussed upgrade of the parameters of the Iachello mass formula to the new PDG data \cite{PDG} on $q \bar q$ mesons.
We set the energy scale,i. e. we determine $\alpha $, by applying Equation (\ref{eq:tetrascorrelati}) to a well-known candidate tetraquark, $a_{0}(980)$, see Ref.\cite{Santopinto:2006my}. Thus, the predicted masses of the other mesons belonging to the same tetraquark nonet are $M(\kappa(800))=0.726\;GeV$, $M(f_{0}(600))=0.354\;GeV$ and $M(f_{0}(980))=0.984\;GeV$ .
These values are in agreement with the experimental values reported in Table 1, even if, before reaching any conclusion, new experiments are mandatory.

\end{document}